\documentclass[prb,aps,twocolumn,epsfig,eqsecnum]{revtex4}
\usepackage{epsfig}
\usepackage{wasysym}
\usepackage{bm}
\usepackage{pstricks}
\newlength{\myL}

\newcommand{\beq}{\begin{equation}}
\newcommand{\eeq}{\end{equation}}
\newcommand{\bea}{\begin{eqnarray}}
\newcommand{\eea}{\end{eqnarray}}

\newcommand{\etal}{{\em et al.}}

\def\tit#1#2#3#4#5{{#1}{\bf #2}, #3 (#4)}
\def\jmp{J.\ Math.\ Phys.\ }

\def\prl{Phys.\ Rev.\ Lett.\ }
\def\pr{Phys.\ Rev.\ }
\def\prb{Phys.\ Rev.\ B\ }

\def\jpco{J.\ Phys.\ Cond.\ Mat.\ }

\def\jpsj{J.\ Phys.\ Soc.\ Jpn.\ }
\def\sci{Science\ }
\def\natu{Nature\ }

\def\jsp{J.\ Stat.\ Phys.\ }

\def\cjp{Can.\ J. Phys.\ }

\def \r {{\bf r}}

\begin{document}


\title{Theory of the [111] magnetization plateau in spin ice}

\author{R. Moessner$^1$ and S. L. Sondhi$^2$}

\affiliation{$^1$Laboratoire de Physique Th\'eorique de l'Ecole Normale
Sup\'erieure, CNRS-UMR8549, Paris, France}

\affiliation{$^2$Department of Physics, Princeton University,
Princeton, NJ 08544, USA}

\date{\today}

\begin{abstract}
The application of a magnetic field along the [111] direction in the
spin ice compounds leads to two magnetization plateaux, in the first
of which the ground state entropy is reduced but still remains extensive. 
We observe that under reasonable assumptions, the remaining degrees of
freedom in the low field plateau live on decoupled kagome planes, and
can be mapped to hard core dimers on a honeycomb lattice.
The resulting two dimensional state is critical, and we have obtained its
residual entropy -- in good agreement with a recent experiments -- the
equal time spin correlations as well as a theory for the dynamical
spin correlations. Small tilts of the field are predicted to lead a
vanishing of the entropy and the termination of the critical phase by
a Kasteleyn transition characterized by highly anisotropic
scaling. We discuss the thermally excited defects that terminate the
plateau either end, among them an exotic string defect which restores
three dimensionality.
\end{abstract}


\maketitle

\section{Introduction}
\begin{figure}
\begin{center}
\epsfig{file=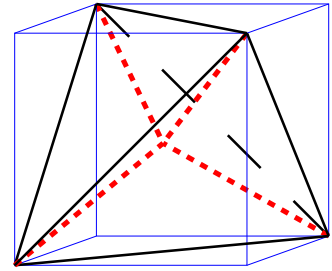,width=0.5\columnwidth}
\end{center}
\caption{ A single tetrahedron inscribed in a cube. In the pyrochlore
lattice, the spins reside on the corners of the tetrahedra. In spin
ice, they are constrained to point along the body diagonals,
${\mathbf{\hat{d}}}_\kappa$, 
indicated by the short-dashed lines. The body
diagonals define the $\left<111\right>$ directions, the cube edges
the $\left<100\right>$ directions, and the bonds the  $\left<110\right>$
directions. }
\label{fig:singleax}
\end{figure}

\begin{figure}
\begin{center}
\epsfig{file=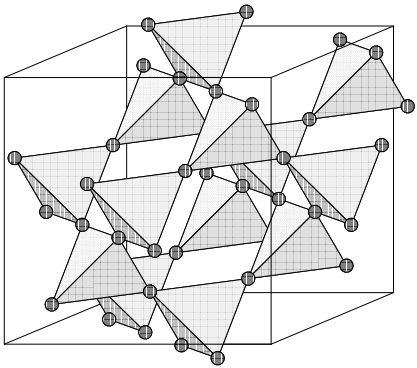,width=0.8\columnwidth}
\end{center}
\vspace{-3cm}
\caption{
The pyrochlore lattice. The cube's axes are the same as those
in Fig.~\ref{fig:singleax}.
}
\label{fig:pyrochlore}
\end{figure}

The discovery of the spin ice compounds
Ho$_2$Ti$_2$O$_7$\cite{harspinice} and
Dy$_2$Ti$_2$O$_7$\cite{ramspinice} is one of the more remarkable events 
in the study of frustrated magnetism in the last decade. 
The name spin ice advertises their statistical mechanics at low temperature,
which can -- approximately -- be mapped onto that of an Ising 
antiferromagnet on the
pyrochlore lattice, which in turn is equivalent to cubic
ice.\cite{andersonpyro} The initial discovery stemmed from the
observation that the large spins $J_{\text{Ho}}=8$ in
Ho$_2$Ti$_2$O$_7$ failed to order at any temperature despite a
ferromagnetic Curie constant.\cite{harspinice} This was understood to
result from the interplay of strong easy-axis single-ion anisotropy
and the geometry of the pyrochlore lattice, which together effectively
turn the ferromagnetic interaction into an antiferromagnetic exchange
between Ising pseudospins -- which describe whether the moment on a
given site is oriented inwards or outwards along the local easy axis
passing through the site and the neighboring tetrahedra (see
Figs.~\ref{fig:singleax} and ~\ref{fig:pyrochlore}).\cite{icebh} 
Later, it was pointed out that
the effective nearest-neighbor ferromagnetic exchange was in large
part due to the effect of dipolar interactions projected onto the
manifold of Ising states.\cite{dipolerahul,dipolebyron}

The antiferromagnetic interaction between the Ising pseudospins
generates an ``ice rule'' -- a minimum energy configuration
must involve two up and two down pseudospins on each tetrahedron.
This ice rule does not, exactly as its cousin the Bernal-Fowler
ice rule does not in the case of crystalline
water, determine a unique ground state. Rather, there
remains a residual extensive zero point entropy, which has been
experimentally observed in the case of the Dysprosium
($J_{\text{Dy}}=15/2$ with $g\mu_B J_{\text{Dy}}\approx 10\mu_B$) spin
ice compound in good agreement with calculations and measurements
of the entropy of ice.\cite{ramspinice} Spin ice therefore offers a
laboratory for studying the properties of water ice by
proxy, but its
properties are, of course, worth studying in their own right. For a review of
this burgeoning field, see Ref.~\onlinecite{ginbrarev}.

It was realized early on that a magnetic field provides a versatile
probe of spin ice, as an external field couples to the actual spin
magnetic moments 
and thus acts on the Ising pseudospins in non-trivial 
ways.\cite{chacha,icerm}
The phenomena predicted here include plateaux in the magnetization 
\cite{liquidgas,siddorder} 
and a liquid-gas transition for fields of different strengths 
and orientations.\cite{liquidgas}

Experimentally, the usefulness of magnetic fields was initially limited by
the absence of single crystals, so that the behavior in a magnetic
field had to be interpreted in terms of an average over all possible
relative angles of fields and crystallites. With the advent of single
crystals, this shortcoming is being 
removed.\cite{bram01,fenn01,gard02,petr02,matsuice} In
recent experiments on Dy$_2$Ti$_2$O$_7$\,\cite{matsuice,hiroiice} 
it was demonstrated that applying a
field in a [111] direction does indeed lead to the predicted pair of
magnetization plateaux---a low field plateau which retains an 
extensive zero temperature entropy albeit one reduced from the zero 
field value, and a second plateau at higher fields where the entropy 
vanishes and the magnetization is saturated upon violation of the 
ice rule.\cite{liquidgas,siddorder}

In this paper we mostly provide a theory of the properties of the 
low field [111] plateau in the $T \rightarrow 0$ limit with some
additional considerations on finite temperature corrections and the
crossovers out of the plateau at low and high fields.
We do so within the nearest neighbor antiferromagnetic model of spin ice
wherein the low temperature limit serves to enforce the ice rule upon
the allowed states. Further, the presence of the magnetic field effects
a dimensional reduction in the same limit---the fluctuating degrees
of freedom are forced to live on decoupled planar subsets of the parent
three dimensional pyrochlore lattice which have the connectivity of
the kagome lattice.
Via a mapping derived previously by us in a study of frustrated Ising 
models in magnetic fields,\cite{isingquant,ogata}
the remaining planar problem maps onto a hardcore dimer model on the hexagonal
lattice. This allows a calculation of the equal time correlations\cite{yokoi86}
-- which are two dimensionally {\em critical} -- and of the (reduced) entropy 
of this region which agrees well with the experiment. We next consider
tilting the field weakly away from the [111] direction, and find that the
system remains in an extended critical phase with
a continuously drifting wavevector,\cite{yokoi86} until it finally
undergoes a continuous phase transition,
known as the Kasteleyn transition in the dimer 
literature,\cite{kasteleyn} where the entropy vanishes.
This transition has a number of interesting features, including the absence of
any symmetry breaking, a mixed first/second order nature and anisotropic
critical exponents. 
The dimer model has a height representation, and as discussed by
Henley,\cite{henleyjsp} this leads to a natural Langevin dynamics
for the coarse grained heights.
We use this to write down expressions for the dynamic spin correlations
in the plateau, which exhibit a dynamical exponent $z_d=2$,
although testing them is likely to be complicated by equilibration 
problems that do not affect the thermodynamics
and statics. Finally, we identify the excitations out of the ground state
manifold, which are a planar zero dimensional object whose condensation
leads to the high field saturated plateau and an unusual infinite string defect,
which restores three dimensionality at low fields and analyze their impact on 
the physics at low temperatures.

In the balance of the paper we will provide details of these assertions.
We begin in Section II by recapitulating the justification for using
the nearest neighbor model and the ice rule and how they give rise
to the plateaux of interest upon addition of a field in the [111]
direction.  We turn next
to the thermodynamics and statics (Section III) and dynamics (Section
IV) of the plateau. We then discuss the complications produced by
the freezing that takes place at low temperatures, in particular
with respect to entropy measurements (Section V),
and then to the impact of thermally excited defects and the longer
ranged dipolar physics in Section VI.  We conclude with a summary. 

\section{The model}

It is not immediately apparent that the spin ice compounds will
exhibit a macroscopic low temperature entropy in zero field, let
alone in a field. Indeed, a sufficiently general microscopic
model for the spin ice compounds involves exchange couplings,
dipolar interactions and a strong easy axis anisotropy.
These can be encapsulated in the classical Hamiltonian for
unit-length spins ${\bf S}_i$:\cite{dipolerahul,dipolebyron}
\bea
{\cal H}&=&
\sum_{(ij)}J_{ij}{\bf{S}}_i \cdot {\bf{S}}_j+
D\sum_{(ij)}\frac{{\bf{S}}_i \cdot {\bf{S}}_j-3
({\bf{S}}_i \cdot {\bf \hat r}_{ij})({\bf{S}}_j \cdot {\bf \hat r}_{ij})
}{|{\bf r}_{ij}|^3} \nonumber\\
&+& {E}\sum_{i}\left({\bf{\hat{d}}}_{\kappa(i)} \cdot
{\bf{S}}_{i} \right)^2 
-g \mu_BJ\sum_i {\bf{B}}\cdot {\bf{S}}_i
\label{eq:hamfield}
\eea
Here $J_{ij}$\ are the exchange constants  and while
the sum on $\left(ij\right)$ runs over all pairs of sites,
only a few are expected to be significant.
The second term is the dipolar interaction of strength $D$, 
where ${\bf r}_{ij}$ is the vector separation of two spins measured in 
units of the nearest neighbor distance, and 
${\bf \hat r}_{ij}= {\bf r}_{ij}/|{\bf r}_{ij}|$. The
third term is the easy axis anisotropy of strength  $E<0$, whose large
magnitude is crucial in these compounds and will be
taken to infinity for the purposes of this paper, thus constraining
the spins to point along their respective easy axes which we have
specified by the unit vectors ${\bf{\hat{d}}}_{\kappa(i)}$ at site
$i$. The
unit cell of the pyrochlore lattice has four sites, which can
be taken to belong to a tetrahedron of one of two orientations, and
hence $\kappa$\ runs from 0 to 3 for the four easy axes that 
point from the center of the tetrahedron to the corner on which 
the site is located (see Fig.~\ref{fig:singleax}). 
These  are the also $\left< 111 \right>$\ directions
of the underlying fcc lattice.  
In the final term, we have allowed for a  magnetic field of strength
$B$ and $g \mu_B J$ is the magnetic dipole moment of the spins.
In the following, we
consider fields along (or close to) the [111] direction. This is a
threefold symmetry axis of the pyrochlore lattice: a field along the
[111] direction singles out the spin ($\kappa=0$) with an easy [111]
axis but leaves intact the symmetry between the other spins (labeled
$\kappa=1,2,3$) with easy axes along the remaining $\left<111\right>$
directions. 

The main difficulty in fixing the parameters in Eq.~\ref{eq:hamfield}
is lack of knowledge of the superexchange, while the value of $D$ can
be essentially fixed via a crystal field calculation. It turns out
that in the spin ice compounds, the effective nearest neighbor exchange is
ferromagnetic by virtue of the dipolar interaction, with the weaker
superexchange possibly being antiferromagnetic and thereby canceling
off part of the dipolar interaction. Very little is known about
further-neighbor superexchange, although there again appears to be a
cancellation effect against the dipolar
interactions.\cite{dipolerahul,dipolebyron}

Despite these uncertainties, both experiment and theory indicate that
a remarkable simplification takes place at moderate temperatures.
If we define the pseudospins $\sigma_i=\pm 1$ by 
whether a spin points into or out of a tetrahedron on a given
sublattice, i.e.\ we write the spins as 
${\bf S}_i=\sigma{\bf {\hat{d}}}_{\kappa(i)}$ then
the accessible low energy states of Eq.~\ref{eq:hamfield} 
in zero field ($\bf B=0$) are largely governed by the ice rule,
which requires that $|\sum_\kappa \sigma_\kappa|=0$ for each
tetrahedron. While Eq.~\ref{eq:hamfield} 
is believed to lead to a unique (up to symmetries)
ground state at $T=0$ in zero field,\cite{dipolerahul,dipolebyron} 
this state has in fact not been observed
experimentally. Provided Eq.~\ref{eq:hamfield} is an appropriate 
description, it thus appears that this state is dynamically inaccessible
and irrelevant to the observed physics.\cite{fn-quantum} 

The net result then is that the accessible behavior is captured by
the greatly simplified nearest neighbor Ising pseudospin Hamiltonian,
\bea
{\cal H}=
J_{\rm eff} \sum_{<ij>} \sigma_i \sigma_j  -g \mu_BJ\sum_i {\bf{B}}\cdot
{\bf {\hat{d}}}_{\kappa(i)} \sigma_i
\ ,
\label{eq:pseudoham} 
\eea
with an antiferromagnetic $J_{\rm eff}$.

The ground states of this Hamiltonian for $B=0$ are, of course, those 
configurations in which $\sum_\kappa \sigma_\kappa=0$ 
(i.e. two spins point in and two out) for each tetrahedron
separately. The number of these states is not known exactly but an
estimate due to Pauling gives ${\cal S}_p/k_B=(1/2)\log(3/2)$ for the
ground state entropy per spin which, as mentioned in the Introduction,
agrees well with the experimental determination of the residual entropy
thus providing support for the simplification.

\subsection{Effect of magnetic field}

The effect of switching on a field is strongly dependent on the direction 
of ${\bf B}$, as first discussed in Ref.~\onlinecite{liquidgas} and is clear from 
Eq.~\ref{eq:pseudoham}.
For instance at zero temperature, an infinitesimal field
along the [100] direction completely lifts the degeneracy of the
ensemble of spin ice ground states while one in the [110] direction
leaves a non-extensive degeneracy. 

A field in the [111] direction, which is our subject in this paper,
orders one sublattice immediately but still leaves a macroscopically 
degenerate set of ground states for a finite range of its values, 
thus producing a magnetization plateau  with a residual zero temperature 
entropy within the ice rule manifold.  At a still higher
field ($g J \mu_B B = 6J_{\rm eff}$) the system abandons the ice rule
and chooses the unique configuration that saturates the magnetic
moment in the [111] direction and thus exhibits a second magnetization
plateau but now with no residual entropy.

To see how this comes about, first note that the projection of the total
spin of a tetrahedron onto the magnetic field is maximized in the case
of $\sigma\equiv-1$ for $\kappa=0$ and $\sigma\equiv1$ for the
others. Hence at sufficiently large fields the system will choose the
unique configuration in which this arrangement holds for all tetrahedra.
This leads, however, to  $|\sum_\kappa \sigma_\kappa|=2$ on
all tetrahedra and is thus in conflict with the ice constraint
$|\sum_\kappa \sigma_\kappa|=0$, so that the low field solution must
be different. Instead in that limit one chooses
$\sigma_\kappa\equiv 1$ for all the spins on sublattice $\kappa=0$ as
their projection onto the external field is maximal but as the other
spins have an equal projection onto the field, one can choose any one
of these to be the second spin with $\sigma=-1$ needed to respect the
ice rule. The transition between these two regimes can be located
by computing the energies of the two arrangements.

The pyrochlore lattice can be thought of an alternate stacking of
kagome and triangular planes, with the triangular planes containing
all the spins of one of the four spin sublattices -- in this case, the
triangular planes of the $\kappa=0$ sublattice are fully polarized and
inert. Consequently, the remaining degrees of freedom live on the
decoupled kagome planes.
Each triangle of a given kagome plane has two spins with a
positive projection ($\sigma=1$) and one with a negative projection
($\sigma=-1$) onto the external field. Such configurations are
equivalent to the ground states of an antiferromagnetic Ising model
($\sigma=\pm1$) with an exchange in excess of the external field
($\sigma=\{-1,1,1\}$ favored over $\sigma=\{-1,-1,1\}$ in each
triangle) \cite{willsice} and as we show in the next section by 
explicit enumeration, they are macroscopic in number.

While we have deduced the low field plateau (henceforth
simply plateau when no confusion is engendered) and its termination
by the saturated state from the nearest neighbor model, its
existence in experiments is further strong evidence
for the applicability of the model and can
be used to deduce the energy scale for the ice rule.\cite{srirrev}

In the next two sections we will analyze the statics, thermodynamics
and dynamics of the plateau at low temperatures within the manifold
of kagome configurations identified above.
In Sect.~\ref{sect:defects}, we will discuss semi-quantitatively the
consequences of the inclusion of thermally excited defects that 
either violate the ice rule or are not confined to the kagome planes.
We also comment briefly there on what might be missed in passing from 
Eq.~\ref{eq:hamfield} to Eq.~\ref{eq:pseudoham} in our problem.

Even with our simplifications we are left with 
a non-trivial statistical and 
dynamical problem that needs to be solved in order to compute the physical 
properties of the plateau and we now turn to this task.

\section{Plateau: Thermodynamics and Statics}

\begin{figure}
\begin{center}
\epsfig{file=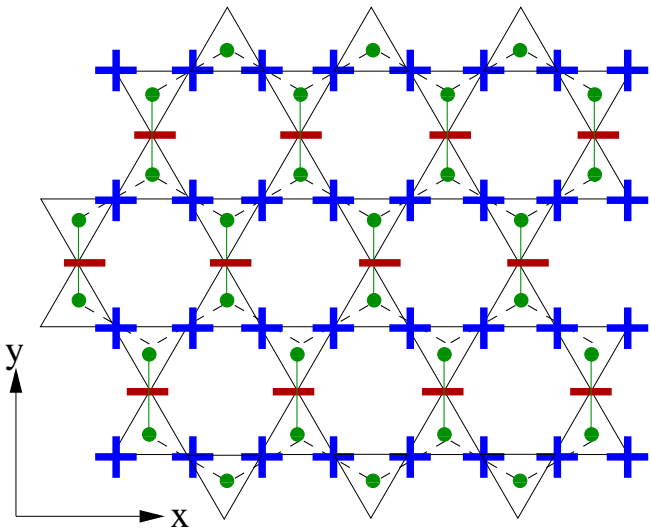,width=1\columnwidth}
\end{center}
\caption{Mapping of pseudospins $\sigma=\pm 1$ on the kagome lattice
onto hardcore dimers on the (dashed) hexagonal lattice. Shown is the
configuration favored by a field tilted slightly
away from the [111] direction. The basis vectors used for
the kagome lattice are also shown.}
\label{fig:icehexmap}
\end{figure}

In the last section we noted that the allowed spin configurations
in a single kagome layer in the plateau are equivalent to the 
ground states of the Ising antiferromagnet on the kagome lattice.
We have previously considered this problem and shown that the ground
states are in correspondence with the configurations of the exactly
soluble problem of the dimer model on the honeycomb 
lattice,\cite{isingquant} a mapping rediscovered by Udagawa
\etal\cite{ogata}
The triangles of the kagome lattice form
a dual hexagonal (honeycomb) lattice, whose bonds are the sites
of the kagome lattice. For each spin with positive projection onto the
field, color in the corresponding link of the hexagonal lattice. As
each triangle has exactly one such spin, each site of the hexagonal
lattice has exactly one colored link emanating from it. By calling
the colored link a dimer, one thus establishes an exact one-to-one
correspondence between the configurations of a hardcore dimer model on
the hexagonal lattice and the spin ice states in a weak [111] field.

\subsection{Entropy}
The entropy of the dimer model on the hexagonal lattice is well known,
having been first computed as the entropy of the equivalent 
triangular lattice Ising antiferromagnet at $T=0$. The latter
has an entropy of ${\cal S}_\triangle
=0.32306 k_B$ per site. This corresponds to an entropy of ${\cal
S}_{\hexagon} ={\cal S}_\triangle/2$ per site of the dimer model. Each
triangle corresponds to a tetrahedron, and hence two sites, of the
pyrochlore lattice, so that the entropy per spin equals
\bea
{\cal S}\approx
0.08077 k_B \ ,
\eea 
which is, of course, also the value obtained in Ref.~\onlinecite{ogata}.

In Ref.~\onlinecite{matsuice}, the value obtained was $0.096\pm0.012
k_B$ per Dysprosium atom. While this work was in progress, another
measurement has appeared, with a value of $0.078 k_B$.\cite{hiroiice}
Our value is just outside the error bars of the former. The fact that
the former is too high suggests that some configurations breaking the
ice rule play a role. Had it been too high, the implication would have
been that a certain degree of (possibly short-range) order, presumably
due to long-range interactions, had already set in. If the latter,
however, should turn out to be the correct value in the end, this
would be an agreement almost too good to have been hoped for.

By comparison, the zero field result of Ref.~\onlinecite{ramspinice}
is ${\cal S}_0\approx0.20 k_B$. This compares to the Pauling estimate
of ${\cal S}_p/k_B= (1/2)\log(3/2)\approx 0.202733$ or the exact value
for two-dimensional spin ice (for which the Pauling estimate is the
same) of ${\cal S}_{Lieb}/k_B=(3/4)
\log(4/3)\approx 0.215762$, so that the decrease due to 
the applied field is by a factor of 2.5 - 2.7.

\subsection{Correlations}

The dimer model describing the plateau has a
range of further interesting features in addition to its nonvanishing
zero point entropy. Most strikingly, its correlations are critical,
decaying as $1/r^2$ at large distances, $r$.

In detail, consider the connected pseudospin correlation function
\bea
c_{\kappa\lambda}({\bf r})=\left<\sigma_\lambda ({\bf r})
\sigma_\kappa(0)\right> -
\left<\sigma_\lambda\right> \left<\sigma_\kappa\right> ,
\eea 
where ${\bf r}$ labels the location of the tetrahedron and the Greek
letters the location of a pseudospin in the tetrahedron. This is
simply related to the correlation functions of the real
spins, $C_{\kappa\lambda}({\bf r})=\left<{\bf S}_\lambda ({\bf r})
{\bf S}_\kappa(0)\right> -
\left<{\bf S}_\lambda\right> \left<{\bf S}_\kappa\right>$. 
For instance, for the components of {\bf S} along the [111] direction,
\bea
C^{[111]}_{\kappa\lambda}= (S/3)^2
(-3)^{\delta_{\kappa,0}+\delta_{\lambda,0}}
c_{\kappa\lambda} .
\label{eq:ps-sp111}
\eea
where the factors of
$3$ are due to the different projections of the inequivalent
easy axes onto the [111] direction. Similarly, the full spin-spin
correlation function is given by 
\bea
C_{\kappa\lambda}=-(S/\sqrt{3})^2
(-3)^{\delta_{\kappa,\lambda}}c_{\kappa\lambda} .
\eea

\begin{figure}
\begin{center}
\epsfig{file=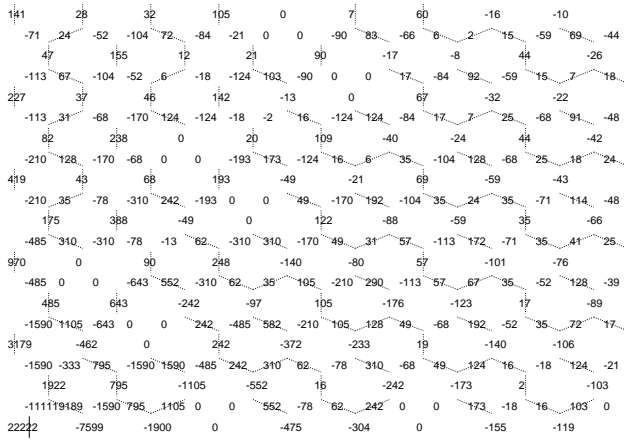,width=1\columnwidth}
\end{center}
\caption{Short distance 
correlations, $10^5\times c_{1\kappa}/4$ of the (pseudo)spin in the
bottom left hand corner, marked by a solid dimer. Positive
correlations are indicated by dashed dimers. This plot uses the same
normalization conventions as that of Tab.~I in
Ref.~\onlinecite{yokoi86}, hence the factor of 1/4; in the convention
of the
dimer model, the correlation at the origin is $-1/9\rightarrow-11111$.
Recall that the dimers occupy the links of the hexagonal lattice,
the midpoints of which are the kagome lattice sites. 
}
\label{fig:shdist}
\end{figure}

In the plateau region, $\sigma_0=\left<\sigma_0\right>=-1$ everywhere,
so that $c_{0,\lambda}\equiv0$. The nontrivial correlations involve
only $\kappa>0$, that is to say spins in the same kagome
planes. These correlations can be calculated following
Ref.~\onlinecite{yokoi86}. We have tabulated the short distance
correlations in Fig.~\ref{fig:shdist}.

The correlations decay algebraically at long distances.
The two independent correlators are,
\bea
c_{11}({\bf r})& \sim &\frac{1}{2\pi^2 r^2}\left[
\cos(4\pi x/3) - \cos(2\theta)
\right] \\ \nonumber
c_{12}({\bf r})& \sim &\frac{1}{2\pi^2 r^2}\left[
\cos(4\pi x/3+4\pi/3) - \cos(2\theta+4\pi/3)
\right] 
\ .
\label{eq:ldcorr}
\eea
Here, ${\bf r}$ is a Euclidean coordinate vector for the kagome
lattice, with $r=|{\bf r}|$ being the distance between two triangles
of the kagome lattice, and $\tan\theta=y/x$, see
Fig.~\ref{fig:icehexmap}. 
This asymptotic behavior, involving a sum
of oscillations at wavevector $q_x = 4 \pi /3$ and a dipolar
piece, can be readily obtained by means of the height representation formulae 
listed in the next section as well.

As a consequence of the first term in brackets in Eq.~\ref{eq:ldcorr},
one would therefore expect a peak in the Fourier transform of the
structure factor at wavevector $\pm(4 \pi/3,0)$. Here we have used
the lattice constant, twice the pyrochlore nearest neighbor distance,
as the unit of length (see Fig.~\ref{fig:icehexmap}). 

\begin{figure}
\begin{center}
\epsfig{file=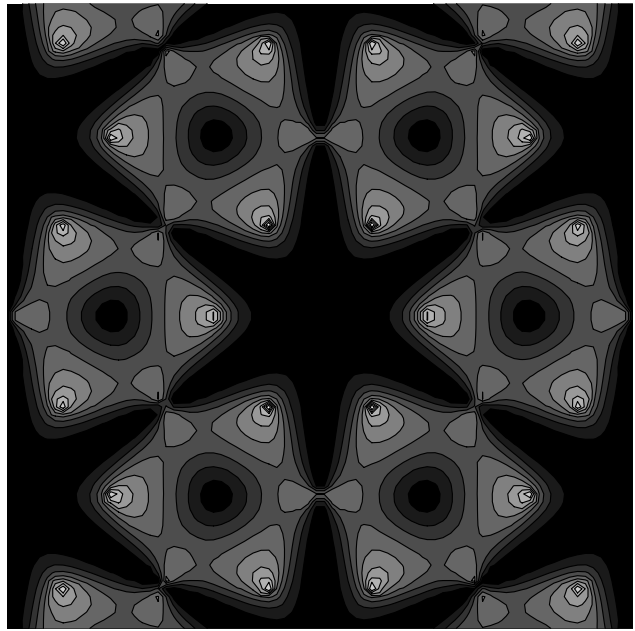,width=1\columnwidth}
\end{center}
\caption{The Fourier transform of the pseudospin
correlations, $c$, in the kagome planes, obtained from a finite system
containing 9604 sites. $q_x, q_y$ range from $-4\pi$ to $4\pi$. In
addition, there is a peak at $q=0$ and the reciprocal lattice vectors
due to the finite average moment induced by the field. Note the
logarithmic peak at $(4\pi/3,0)$ and the symmetry related positions.
Together, they should describe the differential cross section found
in polarized neutron scattering with the neutron spin pointing
along the [111] direction.
Light regions denote strong scattering.
}
\label{fig:strfactor}
\end{figure}

\begin{figure}
\begin{center}
\epsfig{file=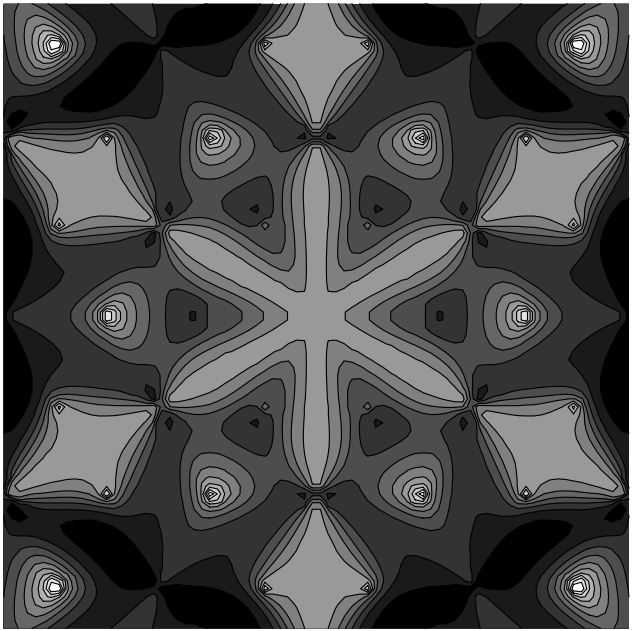,width=1\columnwidth}
\end{center}
\caption{The Fourier transform 
of the correlation of the spins components 
perpendicular to the in-plane wavevector.
Details as in the previous figure. The quantity plotted here is also
the differential neutron scattering cross section for unpolarized
neutrons.}
\label{fig:neutscat}
\end{figure}

The corresponding peaks at the four symmetry related locations are
obtained by the appropriate addition of reciprocal lattice vectors,
$2\pi(1,-1/\sqrt{3})$ and $2\pi(0,2/\sqrt{3})$. Note in particular
that $2\pi(2,0)$ is the reciprocal lattice vector relating the peaks
at $-(4\pi/3) \hat{x}$ and $(8\pi/3)\hat{x}$. However, in
Fig.~\ref{fig:strfactor}, the peak at the latter location is
absent. This happens because the `form factor' of the unit cell has a
zero at $(8\pi/3)\hat{x}$, as can be verified directly from
Eq.~\ref{eq:heightcorr}. In Fig.~\ref{fig:neutscat}, this effect is
reversed in that the peak at $(8\pi/3)\hat{x}$ is the stronger
one; the peak at $(4\pi/3)\hat{x}$, although present, is not visible
on the contour plot for the system size considered as it is almost an 
order of magnitude weaker.

These are not true
Bragg peaks, as there is no long range order. Indeed, as the power law 
decay of the pseudospin correlations 
is rather rapid, $r^{-2}$, their intensity grows only logarithmically
with the planar system size. Similarly, the intensity decreases logarithmically
as one moves away from the center of the peak.  The second term,
although of equal amplitude, does not lead to a feature with 
macroscopic intensity, as no finite fraction of its weight is
concentrated on any one wavevector. In Fig.~\ref{fig:strfactor} we
plot the absolute value of the Fourier transform of the full pseudospin
correlation function, which exhibits these features and is detectable
by polarized neutron scattering. In Fig.~\ref{fig:neutscat} we plot
the cross-section for unpolarized neutrons; the difference in the
two figures reflects the non-trivial relation between the spins and
the pseudospins.
Both figures omit the magnetic Bragg peaks that will arise from the
static magnetization produced by the applied field, and are obtained
for zero out-of-plane wavevector transfer.

\subsection{Kasteleyn transition in a tilted field}

A broader view of the critical correlations in the honeycomb
dimer model is obtained by generalizing it to allow for
unequal fugacities for dimers of different orientations.
As shown by Kasteleyn,\cite{kasteleyn} the equal fugacity
point sits in a critical phase which borders a ``frozen''
phase with 
vanishing entropy that is reached by an unusual transition
that  bears his name. If $z_1$, $z_2$ and $z_3$ are
the fugacities of the three sets of dimers, the transition
takes place when the fugacity of one set equals the
sum of the other two, say $z_1 = z_2 + z_3$. For 
$z_1 > z_2 + z_3$ a unique configuration survives (shown
in Figure 1). It is interesting 
to ask whether this phase transition can be realized in the
spin ice problem. It turns out that this can be done
rather simply by tilting the field.

To see this consider tilting the applied field away from the 
[111] direction so
that it acquires an enhanced component in the [-1-11] direction, 
which is the
easy axis of sublattice $\kappa=1$: ${\bf B}=B \left(\cos\phi
[111]/\sqrt{3}+\sin\phi\right.$ [-1-12]$\left./\sqrt{6}\right)$, so
that the angle the field makes with the [111] direction is given by
$\phi$. This keeps the other two of the three kagome spin sublattices
($\kappa=2,3$) equivalent and singles out the $\kappa=1$
sublattice. To leading order in the tilt angle, spins on sublattice
$\kappa=0$ do not experience a change in energy, whereas 
spins on the other sublattices do:
\bea
E_0^B&=&g\mu_BB J\sigma_0\cos\phi
\nonumber\\
E_1^B&=&-(g\mu_BB J/3)
\sigma_1\left[\cos\phi-2\sqrt{2}\sin\phi\right]
\\
E_{2,3}^B&=&-(g\mu_BB J/3)
\sigma_{2,3}\left[\cos\phi+\sqrt{2}\sin\phi\right] 
\ .
\nonumber
\eea
As the dimer fugacities are $z_\kappa=\exp[2E_\kappa^B/(k_B T)]$,
it follows that the effect of the tilted field is to make them
unequal -- specifically, to privilege the occupation of vertical
dimers over the other two orientations in Fig.~\ref{fig:icehexmap}.
At zero temperature $z_1$ is infinitely bigger than $z_2$ or
$z_3$ at any tilt angle and the system is deep in the frozen phase, 
which is to say the energy gain is all there is and we obtain just 
the so-called staggered 
configuration shown in Fig.~\ref{fig:icehexmap}.

At nonzero temperatures, or finite fugacities, however, the gain in 
energy must compete with the loss of entropy, both extensive, to effect 
a gain in free energy and we obtain a finite range of stability
for the critical phase terminated by the Kasteleyn transition. From
the criterion $z_1=z_2+z_3$ we can deduce a critical tilt angle
$\phi_c$, set by $k_B T=({2\sqrt{2}}/{\ln2})g \mu_B B J\sin\phi$,
at which the transition occurs. Note that the transition temperature
is proportional to the in-plane field strength, $B \sin\phi$, so that 
the experiment can, in principle, be done at $T\ll B$ and when the tilt
angle is sufficiently small to justify our neglect of $O(\phi^2)$ terms.
In the following, we express the dependence on the various parameters 
via $z=z_2/z_1$.

\begin{figure}
\begin{center}
\epsfig{file=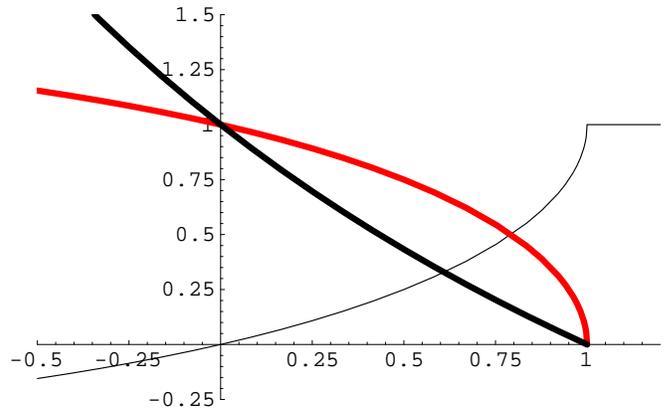,width=1\columnwidth}
\end{center}
\caption{Magnetization of the spins in the kagome planes 
in the [-1-12] direction (thin line) and inverse correlation lengths
(thick lines)
in the $x$ direction and $y$ direction (in black). 
The former is normalized with respect to the saturation
magnetization for $B>B_c$, $m_{{\rm sat}}=
(4\sqrt{2}/3) g \mu_B J$. Saturation for
$B\rightarrow-\infty$ is half this value (and negative).  The inverse
correlation lengths are normalized to their zero field value of
$2\pi/3$. Note that they vanish with different powers at the
transition. 
The $x$ coordinate is given by $(2\sqrt{2}/\ln 2)g \mu_B B
J\sin\phi/(k_B T)$, so that the critical point is located at 1.  }
\label{fig:magcorr}
\end{figure}

Various predictions follow from this analysis:\\ 
(a) The Kasteleyn
transition involves a critical vanishing of the entropy
\bea
{\cal S} \sim (\phi_c-\phi)^{1/2}
\eea
that can be detected via standard thermodynamic measurements.
In equilibrium this implies a significant signature in the tilt
specific heat, $C$, in the form of a divergence,
\bea
C \sim {\partial {\cal S} \over \partial \phi}
\sim(\phi_c-\phi)^{-1/2}
\eea
but freezing is likely to complicate such a direct measurement
as we discuss in Section V.
\\
(b) The expectation values of the Ising spins for $z>1/2$ is given by
\bea
\left<\sigma_1\right>&=&-1+\frac{4}{\pi}
\arcsin[\sqrt{1-\frac{1}{4z^2}}] \\
\left<\sigma_2\right>=\left<\sigma_3\right>&=&
(1-\left<\sigma_1\right>)/2\nonumber \ .
\eea
The magnetization in the [-1-12] direction, $m_\perp$, being
proportional to $\left<\sigma_1\right>$, it follows that it deviates
in the critical region from its saturation value, $m_\perp^{{\rm
sat}}$, as
\bea
m_\perp-m_\perp^{{\rm sat}}\sim(\phi_c-\phi)^{1/2}\ .
\eea
This expression holds to the left of the critical point ($z\geq1/2$,
see Fig.~\ref{fig:magcorr}). To the right, there are no fluctuations,
and
$\left<\sigma_2\right>=\left<\sigma_3\right>=-\left<\sigma_1\right>=1$

The correlations remain critical but change continuously as $B$ 
is tilted. For example, the
equation for the same sublattice connected correlations, 
Eq.~\ref{eq:ldcorr}, is generalized to\cite{yokoi86,fn-sign}
\bea
c_{11}({\bf r^\prime})=\frac{1}{2\pi^2{r^\prime}^2}\left[
\cos(2x/\xi_x) - \cos(2\theta^\prime)
\right]\ .
\label{eq:ldcorrgen}
\eea
Here, ${r^\prime}^2=x^2+(\xi_x/\xi_y)^2y^2$, with
\bea
1/\xi_x&=&2\arcsin\sqrt{1-1/4z^2}\\
1/\xi_y&=&(4z/\sqrt{3})\sqrt{1-1/4z^2}\arcsin\sqrt{1-1/4z^2}\nonumber\ ,
\eea
with $z=z_2/z_1=\exp[2(E_2^B-E_1^B)/(k_B T)]$ and
$\tan\theta^\prime=(x/\xi_x)/(y/\xi_y)$. From this we
observe that:\\
(c) The location of the peak in the structure
factor, which remains logarithmic, is given by $\pm(2/\xi_x,0)$, 
so that it drifts continuously from $(4 \pi/ 3) \hat{x}$
to the center of the Brillouin zone, which it reaches at the phase
transition. Observation of this drift with field tilt should
be a good flag of the unusual critical phase.\\
(d) The scattering pattern is reduced in symmetry -- the applied 
field reduces the six-fold rotational symmetry of the lattice to a 
two-fold one. In particular, this leads to  anisotropic scaling at
the Kasteleyn transition in which there are two diverging correlations
lengths along ($\xi_x\sim (\phi-_c\phi)^{-1/2}$) and transverse 
($\xi_y\sim (\phi_c-\phi)^{-1}$) 
to the in-plane field, whose ratio $\xi_y/\xi_x$ also diverges as 
one approaches the transition, $z\rightarrow 1/2^+$.\\
(e) Finally we note that the transition is asymmetric. 
On the side $z\rightarrow
1/2^-$, no fluctuations are present, so that the transition has an
asymmetric first/second order appearance.
However, the latter property is strictly dependent on the hardcore condition 
on the dimers and tetrahedra violating the ice rule will allow some
fluctuations even beyond the transition, see Sect.~\ref{sect:defects}.

\section{Plateau: dynamics}

We now turn to the dynamical correlations in the plateau 
continuing to assume that the system explores only its ground
state manifold; we will return to the validity of this approximation
in Section V. 
{\it Prima facie}, finding the time dependent
correlations seems a difficult task since the configurations
are characterized by a local constraint, which we have compactly
represented by the hard core dimer mapping. Nevertheless, this can
be done at long wavelengths and low frequencies, following the ideas
of Henley on the dynamical correlations of critical dimer 
models,\cite{henleyjsp} 
which we apply to the honeycomb lattice in the following.
Henley's basic insight is that the dimer configurations on 
bipartite lattices have a height representation whose fluctuations
are {\it unconstrained} at long wavelengths. For the statics this
has been known since the work of Ref.~\onlinecite{Blote82} (see
also Refs.~\onlinecite{zenghenleygenspin,henleyjsp} for a concise
introduction) and the extension to dynamics leads naturally
to a Langevin dynamics for the heights. The resulting theory
is Gaussian and exhibits dynamic scaling with the dynamic exponent
$z_d=2$. We now give brief details of this analysis.

First, we provide a description of the relevant height model. 
Microscopically, this involves a map between
dimer configurations and  the configurations of a surface
specified by giving its local height above the dimer plane.
The microscopic heights are a set of integers, defined on the sites
of the triangular lattice dual to the hexagonal lattice the dimers
reside on. The height changes by +2 $(-2)$ if one crosses a dimer when
going from one site to its nearest clockwise neighbor on an up-
(down-) triangle. If no dimer is crossed, the change is $-1$
(+1). This provides a mapping of dimers onto heights. The dimer
density, $n_d$, is thus given by $n_d=(\nabla^{\rm lat} h+1)/3$, 
where
$\nabla^{\rm lat}$ denotes the lattice derivative corresponding to the rules
defined in this paragraph. 

In the coarse grained, continuum theory, this microscopic expression
indicates the identification,
\bea
n_d= {1 \over 3} (\hat{e}\cdot\nabla) h + {1 \over 3}\ ,
\label{eq:nd}
\eea 
where $\hat{e}$ is a unit vector perpendicular to the 
orientation of the dimer. This is however, not the full expression, even
at leading order.
Upon coarse-graining, a second non-trivial term appears in the expression 
for $n_d$, which reflects the important
fluctuations near the characteristic wavevector of the flat states
-- this is the analog of the staggered ``$2 k_f$'' piece that appears
in the bosonization of one dimensional quantum fermion systems.
This piece can be identified by noting that
the mapping of dimers onto heights is one-to-many: a shift of the
height by 3 units returns the same dimer configuration, and thus the
operator must be invariant under this operation.\cite{Blote82}
One thus obtains for the dimer densities $n_\kappa$:
\begin{widetext}
\bea
n_1-\frac{1}{3}&=&\frac{1}{3}\partial_x h+\zeta \exp(2\pi i h/3) 
\exp(4\pi i x/3) \\
n_2-\frac{1}{3}&=&\frac{1}{3}(-\frac{1}{2}\partial_x+\frac{\sqrt{3}}{2}
\partial_y) h+\zeta \exp(2\pi i h/3) 
\exp(4\pi i x/3+4\pi i/3) \nonumber\\
n_3-\frac{1}{3}&=&\frac{1}{3}(-\frac{1}{2}\partial_x-\frac{\sqrt{3}}{2}
\partial_y) h+\zeta \exp(2\pi i h/3) 
\exp(4\pi i x/3-4\pi i/3) \nonumber .
\eea
where the normalization $\zeta = {1 / (2 \pi a)}$ involves a short 
distance cut
off $a$. 
There are, of course, corrections from less relevant operators which
we have not considered here.

To calculate the static dimer correlators, one uses the fact that the
heights fluctuate in a 
Gaussian manner in equilibrium, \begin{equation} H= \int
d^2 r {K\over 2} |\nabla h|^2 \label{freeE} \end{equation} ($K=\pi/9$
for the honeycomb lattice), whence the height correlator is given as
$\left<h(r)h(0)\right>= -\ln(r/a)/(2\pi K)$. From these we find the asymptotic
correlations,
\bea
c_{ij}({\bf{r}})=\frac{1}{2\pi^2 r^2}\left[
\cos(4\pi x/3+4\pi(j-i)/3) - \cos(2\theta+4\pi(i+j-2)/3)
\right] \ ,
\label{eq:heightcorr}
\eea
\end{widetext}
in agreement with Eqs.~\ref{eq:ldcorr}.
One sees that the two pieces in the dimer correlators arise 
from the ``uniform'' and ``staggered'' pieces of the representations given above.
It is also straightforward to check that the structure factor, at this level
of approximation, gets no contribution from the uniform pieces and consists
entirely of the logarithmic peaks at $\pm {4 \pi/ 3} \hat{x}$ and 
related points. In addition, the extinction of the peaks at 
$\pm {8 \pi/ 3} \hat{x}$ in Fig.~\ref{fig:strfactor} also follows from 
Eq.~\ref{eq:heightcorr}.

To obtain the dynamical correlations, we note that the long wavelength,
low frequency dynamics for a generic local dimer dynamics will be governed
by Henley's  Langevin equation \cite{henleyjsp}
   \begin{equation}
    {{dh(\r)}\over{dt}} = - \Gamma
         {{\delta H }\over{\delta h(\r)}}
               + \zeta (\r,t)
   \label{langevin}
   \end{equation}
where $\Gamma$ is a kinetic coefficient set by microscopics
and the noise $\zeta(\r,t)$ obeys
   \begin{equation}
     \langle \zeta(\r,t) \zeta(\r',t') \rangle = 2 \Gamma
     \delta(\r-\r') \delta (t-t') \ .
   \label{noise}
   \end{equation}
As this is again a Gaussian theory, it follows that the
only non-trivial correlator of the heights is the two-point
function,
\bea
\left<\tilde{h}_q(t)\tilde{h}_{-q}(0)\right>=\frac{1}{K q^2}
\exp[-\lambda(q)t]\ ,
\label{eq:hdyn}
\eea
where $\tilde{h}_q(t)$ is the height configuration at wavevector
$q=(q_x,q_y)$ and time $t$. The relaxation rate for the modes with
wavevectors of magnitude $q$ is given by $\lambda(q)=\Gamma K q^2$,
which implies a critical dynamics with $z_d=2$.

The dynamic correlations can now be obtained from this expression in
the same manner as the static one.  For example, the 
uniform piece of the same sublattice correlator equals 
\bea
\left<{\tilde n_{1_q}}(t)  
{\tilde n_{1_{-q}}}(0)\right>_u=\frac{q_x^2}{Kq^2}\exp[-\lambda(q)t] ,
\eea
which yields the further Fourier transform,
\bea
\left<{\tilde n_{1_{q,\omega}}} 
{\tilde n_{1_{-q,-\omega}}}\right>_u
=\frac{q_x^2}{Kq^2}\frac{q^2}{q^4+w^2} .
\eea
As in the case of static correlations, the structure factor gets no
contribution from such uniform pieces.

The non-zero contribution then comes from the staggered piece which
is first calculated in real space as the vertex operator correlator,
\bea
\langle n_1(r,t) n_1(0,0) \rangle_s &=&
\zeta^2 \{ e^{4 \pi i x/ 3} \langle e^{2 \pi i h(\r)/ 3}
e^{-2 \pi i  h(0)/ 3} \rangle + {\rm c.c.} \} \nonumber \\
&=& 2 \zeta^2  \cos\left(\frac{4 \pi x}{3}\right) 
\exp\left({-\frac{4 \pi^2}{9} 
C(r,t)}\right)\ ,\nonumber
\eea
where
\bea
C(r,t)&=&\left<[h(r,t)-h(0,0)]^2\right>/2=\\
&&\iint \frac{d^2 q}{(2\pi)^2}\frac{1}{K q^2}
[1-\exp(-\Gamma K q^2 t)\cos({\bf q} \cdot \r )] \nonumber \ .
\eea
In the scaling limit, $(r,t) \rightarrow \infty$ with $r^2/t$
fixed, this can be written in the scaling form,
\beq
\langle n_1(r,t) n_1(0,0) \rangle_s = {1 \over 2 \pi^2 r^2} 
\cos\left(4 \pi x \over 3\right) 
 g\left({r^2 \over \Gamma K t}\right)
\label{eq:dyndim}
\eeq
where the scaling function is given in terms of the incomplete Gamma
function as 
\bea
g(x) = e^{-\Gamma(0,{x/ 4})}
\eea 
and exhibits the asymptotics 
\beq
g(x) \sim \left\{
\matrix{ {e^\gamma x/ 4} \hspace{1cm} x \ll 1 \cr 
{4 \exp(-x/4)/ x} \hspace{0.8cm} x \gg1 \cr}
\right.
\ .
\eeq  
The former encodes the autocorrelation 
\bea
\langle n_1(0,t) n_1(0,0) \rangle_s = {e^\gamma \over
8 \pi^2} {1 \over \Gamma K t}\ ,
\eea where $\gamma=0.5772 ...$ is the
Euler-Mascheroni constant.

The remaining task is to obtain the Fourier transform of 
Eq.~\ref{eq:dyndim}
which does not appear possible in closed form and will therefore 
probably
to be accomplished numerically if desired. However, the
essential features can be deduced as follows. 

First,
the Fourier transform will still be peaked about ${\pm {4 \pi/ 3}
\hat{x}}$ and symmetry related points. 
Second, if we measure momenta from each of these values,
the result exhibits the scaling form
\bea
\left<{\tilde n_{1_{q,\omega}}} {\tilde n_{1_{-q,-\omega}}}\right>_s
= {1 \over |\omega|} \tilde{g}\left(
{\Gamma K q^2 \over |\omega|}\right)\ .
\eea
Third,  one can show that 
\bea
\tilde{g}(x) &\sim& \left\{ 
\matrix{ {1/2} \hspace{1cm} x \ll 1 \cr
{c/x} \hspace{1cm} x \gg 1 \cr}
\right.
\eea
with some constant $c$
and that the corrections about either limit are analytic. Together,
the lasts two features imply that fixed frequency cuts will
exhibit peaks of height $({2 |\omega|})^{-1}$, finite with divergent
system size, whose widths will exhibit
the characteristic $z_d=2$ scaling, $\Delta q \sim \sqrt{\omega / 
\Gamma K}$.
The complimentary fixed $q$ cuts will exhibit a diffusive peak
at $\omega=0$ of height ${c/(\Gamma K q^2)}$ and width
$\Delta \omega \sim \Gamma K q^2$.

It is worth noting that in taking the scaling limit we have kept all
information relevant to long wavelengths and low frequencies but
if we attempt to reconstruct the equal time correlator we will
find a spurious ultraviolet singularity. Likewise the large
frequency behavior at a fixed $q$ will be softer than the 
${1/|\omega|}$ dependence implied by the scaling form.

\section{Freezing}

This is a good place to note an important subtlety in making contact
between our analysis, and indeed all theoretical work on ice and spin
ice, and the experimental systems. This is the feature that both ice
and spin ice exhibit diverging relaxation times (set by the temperature
dependent $\Gamma$ in our formalism) at low temperatures
which overtake the timescale of experiments so ergodicity is lost. For
spin ice the evidence for this comes from the experiments of
Refs.~\onlinecite{schifferice,matsuhiraice}, which report a strong
slowdown of the dynamics setting in around 1-2 K, a signature of which
is the appearance of hysteresis in magnetization measurements.
Consequently we need to examine whether the equilibrium
computations of the this paper represent measurable quantities.

The good news is that the thermodynamic and static quantities are
indeed still measurable. For the magnetization and the static
structure factor this is a consequence of self-averaging in the
sample -- with probability one these quantities are the same for
a configuration picked at random as they are for the entire
ensemble of ground states. This in turn comes from two sources. 
First, even in a frozen three dimensional configuration, the different
Kagome planes effectively give different members of the equilibrium
two dimensional ensemble. Second, even in a given plane we get self-averaging.
For example, the spin-spin correlation
function at a fixed separation, averaged over the location of the
spins in a configuration picked at random, converges to its ensemble averaged
value in the limit of infinite system size; the algebraic correlations in our
problem lead to at best a $(\log N)^{1/2}$ correction to the $1/\sqrt{N}$ 
dependence expected for the fluctuations in a system with $N$ sites.
As the structure factor involves exactly this average, all is well on that front. 
The same holds for the magnetization, measured as the moment frozen into
a field cooled sample.

The story with the entropy is different. Indeed it is worth emphasizing
the remarkable fact that experiments measure an entropy associated with
a macroscopic degeneracy of ground states even as the system settles
into just one of them (or a sub-macroscopic number since local fluctuations
presumably do survive even as large scale rearrangements are frozen out).
The contradiction with the statistical mechanical view of entropy
as the logarithmic volume of phase space explored is resolved when one
notes that the experimental determination consists of starting with 
the known entropy of the paramagnetic high temperature state and integrating 
down with the measured heat capacity. At issue then is whether the freezing
substantially affects the ratio of heat capacity to temperature over the 
temperature range where it is significant. For the ice problems, the spectrum
involves a finite gap to making a defect above the ground state manifold.
Consequently, at temperatures below this gap, which is also where freezing
takes place, the heat capacity is exponentially small in the temperature,
whence the freezing hardly affects the entropy determination.\cite{fn-entropy}
In our
problem this implies that field cooled measurements of the heat capacity
will allow determination of the thermodynamic entropy inclusive of tilted
field values.

\section{Thermal and analytic defects}
\label{sect:defects}

Thus far our analysis has assumed that the only accessible configurations
belong to the ground state manifold of the pseudospin Hamiltonian. To
make contact with experiments we need to examine the effects of relaxing
this restriction. In this section we do this, thereby obtaining some insight
into the low and high field boundaries of the plateau and also comment on a 
couple of other salient limitations of our analysis.

As noted earlier, at $T=0$ simple energetics shows that the plateau 
extends over
$0 < g \mu_B J B < 6 J_{\rm eff}$, giving way at zero field to the
full spin ice ground state manifold and to the right to the fully
saturated state. At finite temperatures the plateau
state is no longer field independent but will instead evolve, especially
near the transitions.  
At low temperatures we can gain insight into this evolution
by examining the thermally excited defects that will dress the critical
dimer state that we have discussed in this paper.

\subsection{Monomer defects}

The first defect to consider increases the local magnetization
and it is the condensation of such defects which terminates the plateau
at its high field end. The local minimum energy process to 
consider is one in which a down pseudospin in a kagome plane
is converted to an up pseudospin so that all spins of the two triangles
that share it are now aligned with the field. Such a process violates
the ice rule as there are now two tetrahedra with 
$\sum_\kappa \sigma_\kappa\neq0$, and takes us out of the ground state 
manifold.
A single flipped spin in fact corresponds to a {\em pair} of defects,
which is most easily seen in the dimer representation where it
corresponds to two monomers on adjacent sites of the hexagonal lattice. 
The two partners of the pair can be separated by moving one of the
defects, on an `up' triangle, say, to a neighboring up
triangle. This is done by flipping two spins on an adjacent `down'
triangle, namely the $\sigma=-1$ spin and the spin it shares with the up
triangle. This puts the original up triangle back into the spin ice
ground state at the expense of violating the constraint on the up
triangle sharing the spin with the down triangle. It follows
then that the energy cost of flipping the spin is the creation
energy  $2{\cal E}_m=4 J_{\rm eff}-2 g \mu_B J B/3$ of two
defects. This energy vanishes exactly at critical field 
$g \mu_b J B_c = 6 J_{\rm eff}$ which separates the two plateaux
at $T=0$.\cite{liquidgas,siddorder} 

At finite but low temperatures, the system contains a finite but
small density of these defects whose separation will set a correlation
length and cutoff the critical singularities of the parent dimer
state. Naively, we might anticipate $\xi^2 \sim 1/ n_m 
\sim \exp({{\cal E}_m/k_B T})$
but there is a pseudo-Coulomb (logarithmic) entropic interaction between 
them that modifies this dependence. The exact dependence can be
computed by an energy-entropy balance argument that is equivalent
to a tree level renormalization group computation.\cite{cardybook}
Consider a system of area A and 
let $Z({\bf r_1, r_2})$ be the number of configurations of the dimers
(spin background) in the presence of the two monomers (defects) held
fixed at positions ${\bf r_1}$ and ${\bf r_2}$ while $Z$ is the
number of configurations of the dimers with no monomers present. Then
the free energy cost of introducing two defects is
\bea
\Delta F = 2{\cal E}_m - T \log \int d^2 r_1 \int d^2 r_2 Z({\bf r_1, r_2})/Z
\ .
\eea
The ratio $Z({\bf r_1, r_2})/Z$ can be computed by height representation
theory  by noting that monomers on the two sublattices correspond
to a height mismatch of $\pm 3$ when encircled. The operator identification
described in Ref.~\onlinecite{Blote82} 
then implies 
\bea
Z({\bf r_1, r_2})/Z \sim {1/\sqrt{|{\bf r_1 - r_2}|}}\ ,
\eea
which is the same decay first described in Ref.~\onlinecite{Fisher63}
for the closely related square lattice dimer problem. With this in hand,
it is easy to see that $\Delta F < 0$ when the system size $\xi$, which we
now identify with the correlation length is given by 
\bea
\xi^2 \sim 1/n_m \sim \exp\left({ 8 {\cal E}_m\over7 k_B T}\right)\ .
\eea 

\subsection{Termination of the plateau by monomers}

At a fixed location in the plateau the above formula will describe
the asymptotic low temperature approach to the purely dimer manifold.
At a fixed temperature though this analysis will break down near $B_c$
where a treatment of the statistical mechanics of large numbers of defects
needs to be devised. We expect to address this problem in more detail
elsewhere and here we will content ourselves with three remarks. 

First, matters simplify in a scaling limit $T \rightarrow 0$ and
$B \rightarrow B_c$ with $(B-B_c)/T$ fixed. In this limit we
can ignore all spin configurations save those consisting of dimer 
configurations ``doped'' with some number of monomers. The remaining
problem is the non-interacting monomer-dimer problem and hence the 
interpolation between the two plateaux as a function of $B$ is a 
crossover and not a phase transition.\cite{liebhei} 
Second, at the transition field,
this leads to an equal weight sum over all monomer-dimer 
configurations. The entropy at this 
point is then higher than it is in the low field plateau before it turns 
around and then heads for zero deep into the high field  plateau. 
Third, the transition point exhibits a temperature independent
ensemble in this treatment which should lead to a crossing point for
the magnetization isotherms. Above a critical temperature, the data
\cite{hiroiice} indeed exhibit a maximum in the entropy and a crossing
point for the magnetization isotherms. Below this temperature the
crossover appears to turn into a first order transition at which point
the entropy plummets with temperature and the magnetization develops
a discontinuity.\cite{hiroiice} {\it Prima facie} this appears to be
a puzzle for the nearest neighbor model considered in this paper,
although it is possible that a purely mean field treatment of the
longer ranged pieces of the dipole interaction omitted here renormalize
$B$ sufficiently to turn the sharp low temperature crossover into
a transition.

\subsection{String defects}

The second type of defect to consider is responsible for decreasing
the magnetization towards the low field end of the plateau. As in
this limit we must preserve the ice rule, decreasing the magnetization
requires that we flip a spin on the triangular sublattice $\kappa=0$ 
while satisfying the ice rule by choosing a second spin in the kagome 
plane to have $\sigma =-1$. Interestingly, this is not enough since
the $\kappa=0$ spin is shared by another tetrahedron and so on.
Indeed, one can see quite generally that it must be infinite in length. 
This follows from the
observation that the local ice rule leads to the global property that
all [111] triangular planes have the same magnetization, which is
equal and opposite to that of all kagome [111] planes.\cite{pyroshlo}
As the magnetization of the triangular [111] layers is saturated,
reducing it by flipping one of its spins in one layer requires
flipping one spin in all of the other layers at the same time. The
energy of such a defect, ${\cal E}_s$, is thence most conveniently
quoted per (kagome and triangular bi-) layer. As it involves antialigning
a spin in the triangular and one in the kagome layer with the field,
we have ${\cal E}_{s} = 8 g\mu_B J B /3$.
Remarkably, despite the energy cost proportional to the linear system size,
$L$, it is still entropically favored in a large system.
To see this, note that such a defect corresponds to inserting 
a surplus dimer, violating the hard core condition, into each
kagome plane, which connects a (say) up triangle above which a spin on 
sublattice $\kappa=0$ is flipped with a down triangle below which the next 
flipped $\kappa=0$ spin is located. As in the case of the pair of monomers 
defects, the pair of triangles can again be separated into two
distinct defects---in dimer language into two sites with two dimers
each. If the separation of these sites were to cost no (in plane) entropy, one
would be free to choose which of the $A$ spins in the
triangular layer to flip, thereby endowing the defect with an entropy
of ${\cal S}_2=\ln A$ per layer. For a sufficiently large system,
it would therefore always be free energetically favorable to generate 
such a defect.

The actual density of such defects is lowered by the same in-(kagome)plane
entropic mechanism discussed for monomer defects. Again we appeal to 
height representation theory to find that sites with two dimers carry
charge $\pm 3$ so that the entropic interaction between them is the
same as for two monomers. This implies that per layer the entropic gain
from being able to pick the separation of the defects grows as 
$ \log \int^{\sqrt{A}} r^{-1/2} r dr \sim (3/4)\ln A$. 
From this we deduce that a cylinder of cross sectional area A first
nucleates this string defect when $3/4 \ln A=(8/3)g\mu_B J B$ whence
we expect the area density and hence transverse correlation length
set by  
\bea
\xi^2 \sim 1/n_{s} \sim  \exp[32 g\mu_B J B/9 k_B T]
\eea
at low temperatures. This exponential dependence will then determine
the approach of the magnetization to its plateau value at a fixed
low field as temperature is lowered.

Again, the proliferation of such defects at
low fields but fixed temperature requires a different treatment,
involving a linear response calculation about the full spin ice
manifold, which we will discuss elsewhere. In this regime
all relevant energies are set by the field so that physical quantities 
will be functions of $B/T$ alone. We expect then that the magnetization 
curves will collapse with a finite slope at the
origin when plotted as a function of $B/T$.

We can draw one further inference from our computation of the defect
densities. By equating the activation energies of the two defects
we can identify the field at which their densities cross at the lowest
temperatures---this will also be the field at which the magnetization 
isotherm crosses the zero field value of the magnetization at low
temperatures and hence a second crossing point. This yields a field 
$g \mu_B J B  = (18/31) J_{\rm eff}$
which is about a tenth of the critical field between the plateaux.

How does the presence of such defects alter the results we have
described above? Fundamentally, their presence 
will of course make itself known  as a deviation from the `exact'
result;
in particular, the smallest of the defect induced finite
correlation lengths will determine the cutoff at which, for example,
the logarithmic peaks in the neutron scattering stop growing.

As for the Kasteleyn transition, both types of defects will
inevitably smear out the fluctuation-free regime and therefore the
mixed first/second order nature of the transition. Monomer defects can
be exponentially suppressed by lowering the temperature (compared to
$J$).  As one lowers the temperature at small fields $B/T\ll1$, the
angle $\sin\phi$ at which the transition takes place decreases
inversely with $B/T$, whereas the density of string defects is
exponentially suppressed. By achieving an improved angular resolution,
the crossover from Kasteleyn behavior to a more conventional second
order phase transition could thus be reduced.

\subsection{Disorder and dipoles}

Finally we turn to two significant limitations of our analysis in this
paper. First, actual samples are likely to contain structural
defects due simply to chemical disorder such as
vacancies or interstitials affecting site occupancy or exchange paths.
We are not aware of a determination of the density of such defects,
although for Heisenberg spins on the related SCGO lattice, there have
been both experimental\cite{schifferdaruka,nmrvac} and
theoretical\cite{theoryvac} attempts to determine the density of
vacancies from thermodynamic\cite{schifferdaruka,theoryvac} or NMR
experiments.\cite{nmrvac} As the chemical defect density in single
crystals tends to be higher than in powder samples, this might be a
not insubstantial effect in this context.

The second important feature omitted from the nearest-neighbor spin
ice model are the effects of the long-range dipolar interactions beyond
the nearest neighbor piece, which are sizeable due to the large spin 
of the Dysprosium ion. We have already alluded to one possible 
effect in our discussion of the transition between the two plateaux,
namely that the polarization of the spins may require a self-consistent
treatment of the field $B$ that acts upon them. While this is always
necessary when a macroscopic magnetization is present, in our case the
issue is somewhat more delicate since the largest piece of the dipolar
interactions has already been accounted for in the nearest neighbor model.

On a fundamental level, however, the long-range dipolar interactions
do not seem to lead to a significant inter-plane ordering effect, as
this would have reduced the entropy determined in the experiment. This
may, however, be a consequence not of the precise thermodynamic
behavior of the spin ice Hamiltonian in a field, but rather an indication of the
magnet's inability to access its true ground state in the presence of
energy barriers as discussed in Section II.

\section{Summary}

The application of a field in the [111] direction to the spin ice
compounds leads, by a reasonable set of approximations, to an elegant
dimensional reduction of the three dimensional problem onto
a set of decoupled two dimensional problems. Fortunately, the
resulting two dimensional problem is one of planer dimers and hence
is exactly soluble, so that the
statics and thermodynamics can be determined exactly. While the
computed entropy has already been measured, the predictions for
the correlations can be tested by scattering. Also testable are
thermodynamic and static predictions for a Kasteleyn transition
upon tilting the field in the [-1-11] direction and for the
dynamic correlation in the plateau. Finally we have sketched
a theory of the finite temperature modifications which we intend
to flesh out in future work.\cite{ms-wip}

From the viewpoint of spin ice physics, it is fortunate that
much existing technology turns out to be especially suited
to this task. From the perspective of statistical mechanics
the realization of the hexagonal dimer model as well as of
the monomer-dimer problem in a three dimensional system with
built in self-averaging and easy access via neutron scattering, 
in contrast to surface or interface realizations, 
is surely interesting.

Sadly it does not appear possible to make one final link---to
the {\em quantum} dimer model on the hexagonal 
lattice\cite{Rokhsar88,sachhex,mschex} as this would require
a ``resonance'' quantum dynamics consisting of a 
simultaneous coherent tunneling of six pseudospins which
is rather unlikely given the large spin $J=15/2$ of the
constituents. We leave the realization of this physics as
a challenge for future work.

\section{Acknowledgements}

We would like to thank Leon Balents for interesting discussions, and
for drawing our attention to Ref.~\onlinecite{matsuice},
to Sriram Shastry for many discussions about spin ice, and to
Premi Chandra for collaboration on closely related work. We
are also grateful to Elliott Lieb and David Huse for enlightening
discussions on measuring the entropy of ice.
RM is grateful to the Aspen Center for Physics and the Lorentz
Centre of Leiden University, where parts of this work were undertaken.
This work was in part supported by the Minist\`ere
de la Recherche et des Nouvelles Technologies with an ACI grant.
SLS would like to acknowledge support by the NSF (DMR-9978074 and 
0213706) and the David and Lucille Packard Foundation.


\begin{thebibliography}{99}
\bibitem{fn-quantum} A different possibility is the selection of
a unique state in the ice manifold by quantum effects, e.g. as 
discussed for two dimensional ice in Ref.~\onlinecite{mst-ice}, 
but that prospect is also irrelevant to observations under
laboratory conditions.
\bibitem{fn-sign} This expression differs by an overall sign from
Eq.~37 of Ref.~\onlinecite{yokoi86}. It combines (in their notation)
$C_{vv}$ and $C_{vv^\prime}$, allowing for half integer values of
$x$. The oscillation period, for a fixed angle, $\theta^\prime$, of
this expression varies as $1/\cos\theta^\prime$, as noted in
Ref.~\onlinecite{yokoi86}. This is a consequence of the
projection of the period of oscillation in the $x$ direction onto the
direction determined by $\theta^\prime$.
\bibitem{fn-entropy} Thus the equality of the measured entropy and its
equilibrium computation is not a consequence of 
self-averaging. Indeed the temperature range where the system freezes
and we need to invoke self averaging contributes nothing to the entropy 
determination since it is characterized by a zero heat capacity (a
temperature independent internal energy).
\end{thebibliography}
\end{document}